# Probing the A1 to $L1_0$ Transformation in FeCuPt Using the First Order Reversal Curve Method

*Dustin A. Gilbert,*[1] *Jung-Wei Liao,*[2] *Liang-Wei Wang,*[2] *June W. Lau,*[3] *Timothy J. Klemmer,*[4] *Jan-Ulrich Thiele,*[4] *Chih-Huang Lai,*[2] *and Kai Liu*[1,*]

[1]Physics Department, University of California, Davis, CA 95616 USA

[2]Department of Materials Science and Engineering, National Tsing Hua University, Hsinchu, Taiwan

[3]National Institute of Standards and Technology, Gaithersburg, Maryland 20899, USA

[4]Seagate Technology, Fremont, CA 94538 USA

Abstract

The $A1$-$L1_0$ phase transformation has been investigated in (001) FeCuPt thin films prepared by atomic-scale multilayer sputtering and rapid thermal annealing (RTA). Traditional x-ray diffraction is not always applicable in generating a true order parameter, due to non-ideal crystallinity of the $A1$ phase. Using the first-order reversal curve (FORC) method, the $A1$ and $L1_0$ phases are deconvoluted into two distinct features in the FORC distribution, whose relative intensities change with the RTA temperature. The $L1_0$ ordering takes place via a nucleation-and-growth mode. A magnetization-based phase fraction is extracted, providing a quantitative measure of the $L1_0$ phase homogeneity.

High magnetic anisotropy materials are critical to key technologies such as ultrahigh density magnetic recording[1] and permanent magnets.[2] Among them, ordered FePt alloys in the $L1_0$ phase are particularly sought after. They are ideal candidate materials for the emerging >1 Terabits/in$^2$ heat-assisted magnetic recording (HAMR) media[3, 4] or bit-patterned media,[5-7] as well as high energy-product permanent magnets,[6, 8, 9] as they possess some of the largest anisotropy values known.[10] However, the highly desirable properties are associated with the tetragonal $L1_0$ phase. A critical challenge has been the elevated annealing temperature (typically 600-650 °C) necessary to transform the as-deposited cubic $A1$ phase into the ordered $L1_0$ phase,[11] which is often incompatible with the rest of the manufacturing processes. Another key issue is the stringent requirement on switching field distribution for ultrahigh density HAMR media (<5%),[10, 12] where any residual $A1$ phase could be detrimental. Therefore, a complete understanding of the low anisotropy $A1$ to high anisotropy $L1_0$ phase transformation is critical. In prior studies, an order parameter $S$ is often used to characterize the degree of the $L1_0$ ordering, using properly corrected, integrated intensities of the x-ray diffraction (XRD) peaks, since differences in lattice occupation are manifested in the structure factor.[13-15] This is an elaborate, time consuming process, yet its applicability still has significant limitations, e.g., for textured samples or materials with poor crystallinity, and for samples with tiny amount of minority phases.

We have recently demonstrated a convenient synthesis route to achieve (001) oriented FeCuPt films, using atomic-scale multilayer sputtering (AMS) and rapid thermal annealing (RTA) at 400 °C for 10 seconds, which is significantly more benign compared to earlier studies.[16] Magnetic properties of these films can be conveniently tuned with Cu content to achieve high anisotropy, large saturation magnetization and moderate Curie temperature, a combination highly desirable for HAMR media. In this work, we have employed the first order reversal curve (FORC) method[17-20] to qualitatively and quantitatively investigate the $A1 - L1_0$ phase transformation, particularly the RTA temperature dependence. We show that the $L1_0$ ordering takes place via a nucleation-and-growth mode during the non-equilibrium synthesis, and the magnetic phase fraction extracted by FORC is a convenient and complementary measure of the $L1_0$ ordering, which directly captures distributions in the relevant magnetic characteristics.

Atomic-scale multilayer films with nominal structure of $[Fe(0.9Å)/Cu(x)/Pt(1.4Å)]_{16}$ were grown by DC magnetron sputtering from elemental targets on amorphous $SiO_2$(200 nm)/Si substrates in a vacuum chamber with base pressure of $7\times10^{-7}$ torr. The Cu thickness ($x$) is varied to adjust the stoichiometric ratio (e.g., 0.4 Å for $Fe_{39}Cu_{16}Pt_{45}$). The introduction of Cu has been shown to selectively replace Fe atoms and help to tune the material properties to be desirable for HAMR applications.[16] Subsequently the films were annealed by RTA at various temperatures (10s rise + 10s dwell time) in $1\times10^{-5}$ torr vacuum using infrared (IR) heating lamp with wavelength of 400–1100nm.[21] Due to its small band gap of 1.1 eV, the Si substrate readily absorbed the IR light, in contrast to the amorphous $SiO_2$, which has a much larger 8.9 eV band gap. The thin FePt was mainly heated through thermal conduction across the $SiO_2$ barrier. As a result, the quick thermal expansion of the Si substrate exerts significant tensile stress on the atomic-scale multilayers of FePt across the $SiO_2$ and transfer the thermal energy to assist the $L1_0$ ordering as well as (001) orientation of the film.[16, 21, 22] After RTA the films were capped with a thin layer of Ti (20 Å) to prevent oxidation. For simplicity the rest of the paper will mainly focus on a series of $Fe_{39}Cu_{16}Pt_{45}$ samples (referred to as FeCuPt hereafter), while results from other samples are provided in the Supplementary Online Material.[23]

X-ray diffraction was performed using a Bruker D8 thin film x-ray diffractometer with Cu K$_\alpha$ radiation, and used for phase identification following procedures outlined previously.[15] Film microstructure and morphology were characterized by electron diffraction, transmission electron microscopy and atomic force microscopy. Magnetic measurements were performed using vibrating sample magnetometry (VSM) at room temperature with the field applied perpendicular to the films, unless otherwise noted. To investigate detailed magnetization reversal, FORC measurements were performed as follows:[17-20] From positive saturation the magnetic field was swept to a reversal field $H_R$, where the magnetization, $M(H,H_R)$, was measured under increasing applied field, $H$, back to positive saturation, tracing out a FORC. The process was repeated for decreasing $H_R$ until negative saturation was reached. The normalized FORC-distribution was then extracted,[24]

$$\rho(H, H_R) \equiv -\frac{1}{2M_S}\frac{\partial^2 M(H,H_R)}{\partial H\, \partial H_R}, \tag{1}$$

where $M_S$ is the saturation magnetization. Alternatively the FORC distribution could be represented in another coordinate system defined by local coercivity $H_C=(H-H_R)/2$ and bias field $H_B=(H+H_R)/2$.

X-ray diffraction (XRD) patterns for the $Fe_{39}Cu_{16}Pt_{45}$ films are shown in Fig. 1. For RTA temperatures $T_{RTA}<350°C$ there are no appreciable FeCuPt peaks, indicating poor initial crystalline ordering. At $T_{RTA} \geq 350°C$, both (001) and (002) peaks are clearly observed. The presence of the (001) peak indicates the establishment of the $L1_0$ phase, as it is forbidden in the $A1$ phase. The order parameter $S$ is calculated from the $I_{(001)}/I_{(002)}$ ratio to quantify the degree of $L1_0$ ordering, where $I_{(001)}$ and $I_{(002)}$ are the integrated intensity of the (001) and (002) peak, respectively, corrected for absorption, Lorentz factor, Debye-Waller factor, and angular dependent atomic scattering factors.[13-15] The order parameter is normalized to the calculated maximum value, $S_{Max}=1-2\Delta$, where $\Delta=0.05$ is the variation in the (FeCu):Pt atomic ratio from 50:50. The $S/S_{Max}$ is determined to be 1.0, 0.97, and 0.97 for $T_{RTA}$ of 350°C, 375°C, and 400°C, respectively, seemingly suggesting a high degree of $L1_0$ ordering at these temperatures.

However, at $T_{RTA} < 350°C$, the absence of (001) or (002) peak makes it impractical to extract the $L1_0$ ordering parameter using this approach. The XRD order parameter is dependent on $(I_{(001)}^{L1o} + I_{(001)}^{A1})/(I_{(002)}^{L1o} + I_{(002)}^{A1})$; the (001) peak is forbidden in the $A1$ phase, $I_{(001)}^{A1} = 0$, thus $S/S_{Max}$ will scale between 0 for the fully $A1$ phase to unity for the fully $L1_0$ phase. In the present case, no (002) peak is observed for the $A1$ phase, thus $I_{(002)}^{A1} = 0$, and $S$ is only dependent on $I_{(001)}^{L1o}/I_{(002)}^{L1o}$, which is identical to $S_{Max}$. Therefore $S$ does not vary with the phase transformation and cannot be used in its usual context to measure the $L1_0$ ordering. More generally, when the sample's initial crystalline ordering is non-ideal (e.g., the (002) intensity changes substantially during annealing), the order parameter $S$ is no longer applicable in gauging the $L1_0$ ordering extent.

Transmission electron microscopy (TEM) and electron diffraction show that all samples have crystalline grains (see Supplementary Online Material[23]); at $T_{RTA}$=325°C, both rounded and elongated grains are observed, with an average size of ~ 6nm; at $T_{RTA}$= 400°C, only elongated grains are observed, with an average size of ~ 12nm, which are consistent with the effects of the tensile stress during the RTA. It is the crystalline nature of the films that allows us to use the conventional $S$ parameter to gauge the degree of ordering. Cross-sectional TEM studies on a similar series of $Fe_{52}Pt_{48}$ samples (i.e., Cu content is zero) illustrate that during the phase transformation, residual disordered $A1$ regions exist only in certain parts of the otherwise (001) oriented $L1_0$ films (Fig. 2), rather than homogeneous dispersion over numerous sites throughout the films. However, it is difficult to quantify the $L1_0$ phase fraction from TEM alone.

Room temperature hysteresis loops in the out-of-plane (OOP) and in-plane (IP) geometries for the $Fe_{39}Cu_{16}Pt_{45}$ samples annealed by RTA at 300-400°C are shown in Fig. 3. After RTA at 300°C and 325°C, the OOP major loops are nearly closed with negligible remanence while the IP loops are sharp with large remanence and small coercivity, indicating an out-of-plane hard axis and suggesting that the films are primarily in the low anisotropy $A1$ phase. As the RTA temperature $T_{RTA}$ is increased to 350°C, the OOP loop shows a gradual increase in remanence and coercivity, still showing largely hard-axis behavior, while the IP loop shows a sudden increase in coercivity and decrease in remanence. A more distinct change is observed for RTA at 375°C and 400°C, where the OOP loop exhibits easy-axis behavior with remanence approaching unity and a much enhanced coercivity,[25] and the IP loop has the smallest remanence, suggesting the film is primarily $L1_0$ ordered with out-of-plane anisotropy. The $T_{RTA}$-dependent remanence, coercivity and saturation magnetization are shown in Figs. 3(c) and 3(d). The continuous changes of the remanence and coercivity illustrate a gradual evolution of the $A1$-$L1_0$ phase transformation with increasing $T_{RTA}$.

To further investigate the $A1$-$L1_0$ phase transformation and the magnetization reversal behavior of the films, FORCs have been measured in the OOP geometry for the $Fe_{39}Cu_{16}Pt_{45}$ samples. For $T_{RTA}$ of 300°C, the family of FORCs is highly compressed inside the slanted major loop [Fig. 4(a)]. The corresponding FORC distribution $\rho(H_C, H_B)$ shows a single vertical ridge located near $H_C=0$ [Figure 4(b)], which is clearly seen in the normalized projection of $\rho(H_C, H_B)$ onto the $H_C$ axis, $(dM/dH_C)'$ [Figure 4(c)]. These characteristics show that the magnetization is largely reversible and the coercivity is small, indicating that the sample is in the low anisotropy $A1$ phase.

For $T_{RTA}$ of 325°C and 350°C, the family of FORCs fills the slanted hard-axis major loop with higher remanence and larger coercivity, indicating an increasing degree of irreversibility [Figures 4(d) and 4(g), respectively]. In the FORC distributions, the vertical ridge near $H_C=0$ is still the most outstanding feature; interestingly, a second horizontal FORC feature located at a higher $H_C$ of ~ 2.8 kOe begins to emerge at 325°C [Figures 4(e) and 4(f)] and becomes quite appreciable at 350°C [Figures 4(h) and 4(i)]. These features demonstrate that the $L1_0$ ordering is beginning to occur, however, the $A1$ phase is still the majority phase that dominates the magnetization reversal.

For $T_{RTA}$ of 375°C and 400°C, the family of FORC's show qualitatively different easy-axis behavior [Figures 4(j) and 4(m)]. In the FORC distribution, the vertical ridge near $H_C=0$ has largely vanished and the horizontal FORC feature has become quite prominent, with its center shifting to higher $H_C$ values [$H_C$= 3.2 kOe in Figures 4(k) and 4(l) for 375°C, and 3.7kOe in Figures 4(n) and 4(o) for 400°C]. These samples exhibit essentially full $L1_0$ ordering and the disappearance of the $A1$ phase.

The two distinct FORC features directly correspond to the low and high anisotropy $A1$ and $L1_0$ phase, respectively. Their coexistence at intermediate $T_{RTA}$ of 325°C and 350°C, while the major loop only shows the hard-axis reversal, demonstrates the capability of FORC in deconvoluting magnetic phases. Furthermore, the evolution of these two FORC features under increasing $T_{RTA}$ reveals details of the $A1$- $L1_0$ phase transformation mechanism. The fact that these two features remains well separated in the FORC distribution (in $H_C$-$H_B$ coordinates), and their relative intensity changes under increasing $T_{RTA}$, is indicative of a nucleation-and-growth mechanism of the $L1_0$ phase.[26] Under this mechanism, annealing changes the relative ratios of the magnetically hard and soft phases, primarily modifying the intensity of the corresponding

FORC features. This is consistent with the TEM image shown in Figure 2, which illustrates certain residual $A1$ "pockets" in the matrix of $L1_0$ films. Alternatively, had the phase transformation with $T_{RTA}$ been realized through a uniform growth mode, we would have expected a gradual shift of a single FORC feature *as a whole*, from low to high coercivity, and more extensive $A1$ regions in the TEM view. Thus the FORC technique is able to uniquely identify and separate contributions from the hard and soft phases, despite the single phase appearance of the major loop.

Furthermore, the integration $I_{Irrev} = \int \rho(H_R, H) dH_R dH = \int \rho(H_B, H_C) dH_B dH_C$ for $H \geq H_R$ captures the total amount of irreversible magnetic switching in a sample.[27] Thus, it can be used to quantitatively determine the amount of magnetic phases.[28, 29] By selectively integrating the normalized FORC distribution (Equation 1) over the horizontal feature corresponding to the $L1_0$ phase, compared to the saturation magnetization of the entire sample, we can extract a magnetization-based $L1_0$ phase fraction. Since the saturation magnetization of the $A1$ (818 emu/cm$^3$) and $L1_0$ (790 emu/cm$^3$) phases are nearly identical, the magnetic phase fraction can be directly correlated to the structural phase fraction.[30, 31] In addition, it is possible to reconstruct the major hysteresis loops of each phase.

The evolution of the $L1_0$ phase fraction in $Fe_{39}Cu_{16}Pt_{45}$ determined by FORC with $T_{RTA}$ is shown in Fig. 5, indicating a gradual increase in the $L1_0$ ordering up to 350°C, followed by an abrupt ordering near 375°C. This is consistent with the kinetic ordering temperature reported earlier.[26, 32] For comparison, the XRD order parameters are also included in Figure 5, which show good agreement only for highly $L1_0$ ordered 400°C sample. Similar trends are also found in another series of $Fe_{28}Cu_{27}Pt_{45}$ films (see Supplemental Material[23]). While the structural order parameter is not always readily available (such as the case here for the 300°C and 325°C samples) or reliable (depending on capturing the proper crystal planes of the $A1$ phase), the magnetization-based phase fraction provides a complementary and direct measure of the $L1_0$ phase homogeneity, which is critical to SFD and eventual application of such materials in HAMR media.

In conclusion, we have investigated the $A1$- $L1_0$ phase transformation in (001) FeCuPt thin films prepared by a non-equilibrium strain-assisted synthesis approach. The $A1$ and $L1_0$ phases manifest themselves as two distinct features in the FORC distribution, whose relative intensities change as the RTA temperature increases from 300°C to 400°C. The $L1_0$ ordering takes place via a nucleation-and-growth mode. Traditional x-ray diffraction suggests an unrealistic order parameter due to poor initial crystalline ordering of the $A1$ phase. An alternative magnetization-based $L1_0$ phase fraction is extracted. The FORC method not only sheds insight into the $L1_0$ ordering mechanism in these ultrathin films, but also provides a quantitative measure of the $L1_0$ phase homogeneity. This approach is applicable to other $L1_0$ materials such as MnAl[33] and MnGa[34] that are being keenly pursued for rare-earth-free permanent magnets, as well as magnetic phase separation studies in general.

This work has been supported by the NSF (DMR-1008791). Work at NTHU has been supported in part by the Hsinchu Science Park of Republic of China under Grant No. 101A16.

**Figure captions**

**Figure 1.** X-ray diffraction patterns of $Fe_{39}Cu_{16}Pt_{45}$ thin films annealed by RTA at 300-400°C. The sharp (002) peak of the Si substrate is suppressed in some of the samples due to a slight deviation from the Bragg condition, which has no appreciable effect on the FeCuPt peaks.

**Figure 2**. Cross-sectional TEM view of an $Fe_{52}Pt_{48}$ film showing a residual disordered $A$1 region, highlighted by dashed lines, embedded in the (001) oriented $L1_0$ film. The film was capped by an $FeO_x$ layer and annealed by RTA at 500 °C, which had comparable $L1_0$ ordering characteristics as the $Fe_{39}Cu_{16}Pt_{45}$ film capped by Ti and annealed by RTA at 400 °C.

**Figure 3.** Room temperature hysteresis loops in the (a) out-of-plane and (b) in-plane geometries of $Fe_{39}Cu_{16}Pt_{45}$ thin films annealed by RTA at 300-400°C. Trends for the (c) coercivity and remanence and (d) saturation magnetization are also included.

**Figure 4.** Family of FORC's (top row), corresponding FORC distributions (middle row) and their normalized projections $(dM/dH_C)$' onto the local coercivity $H_C$ axis (bottom row) for $Fe_{39}Cu_{16}Pt_{45}$ samples annealed at 300° (a-c), 325° (d-f), 350° (g-i), 375° (j-l), and 400°C (m-o).

**Figure 5.** $L1_0$ phase fraction determined by FORC (squares) and the order parameter determined by x-ray diffraction (triangles) for $Fe_{39}Cu_{16}Pt_{45}$. Error bars were determined by the error on the peak area fitting (XRD) and the propagated background standard deviation (FORC).

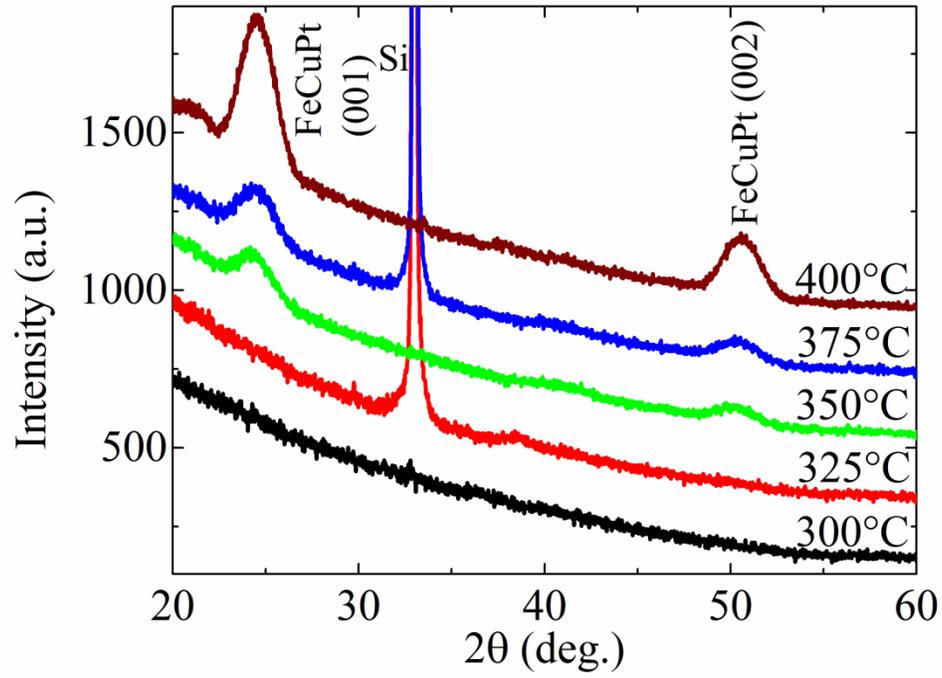

**Fig. 1**

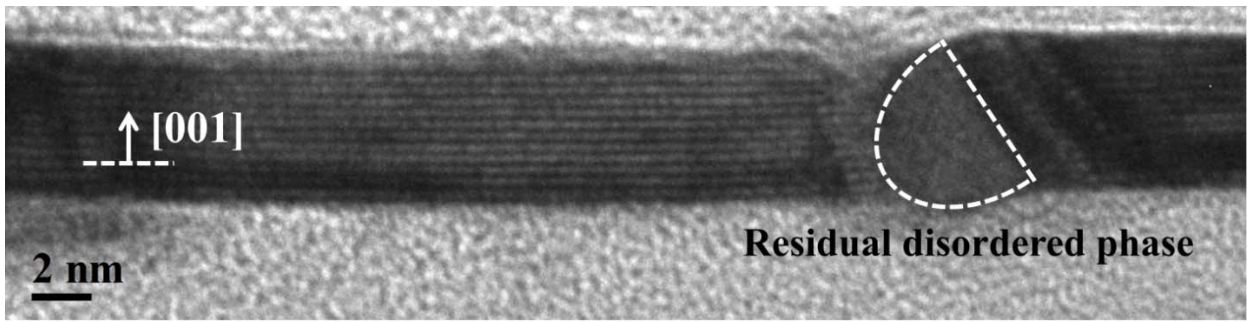

**Fig. 2**

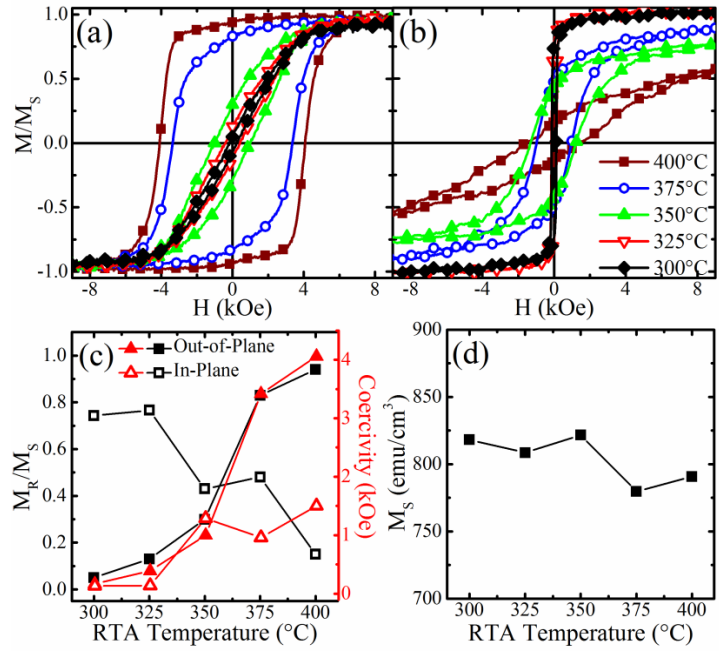

**Fig. 3**

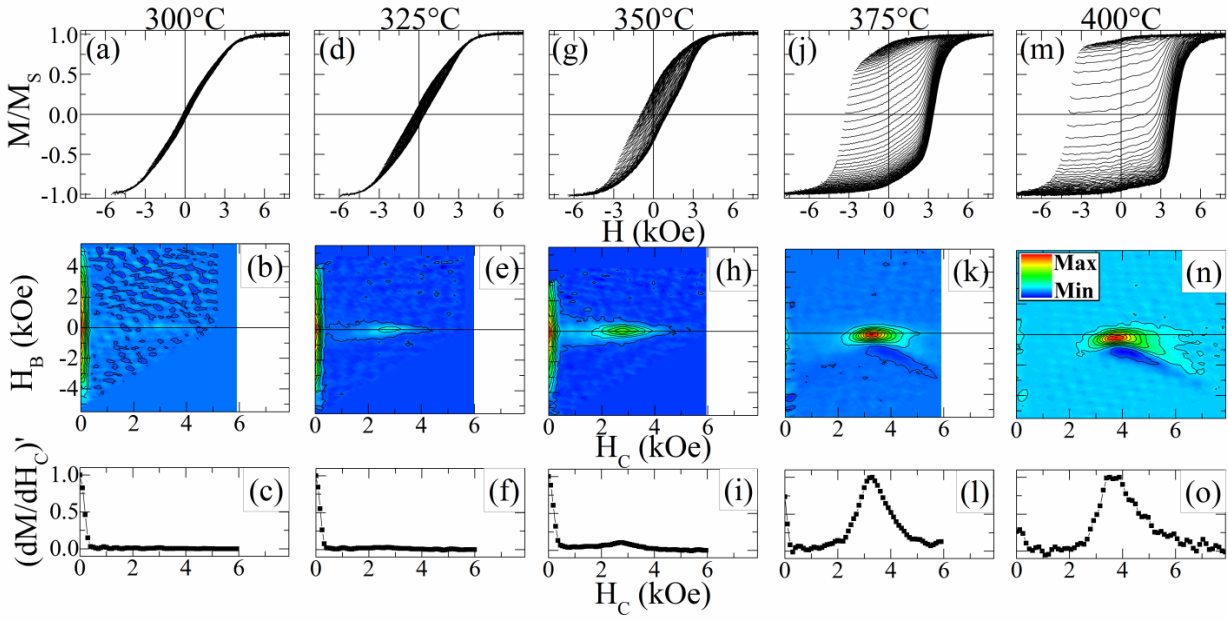

Fig. 4

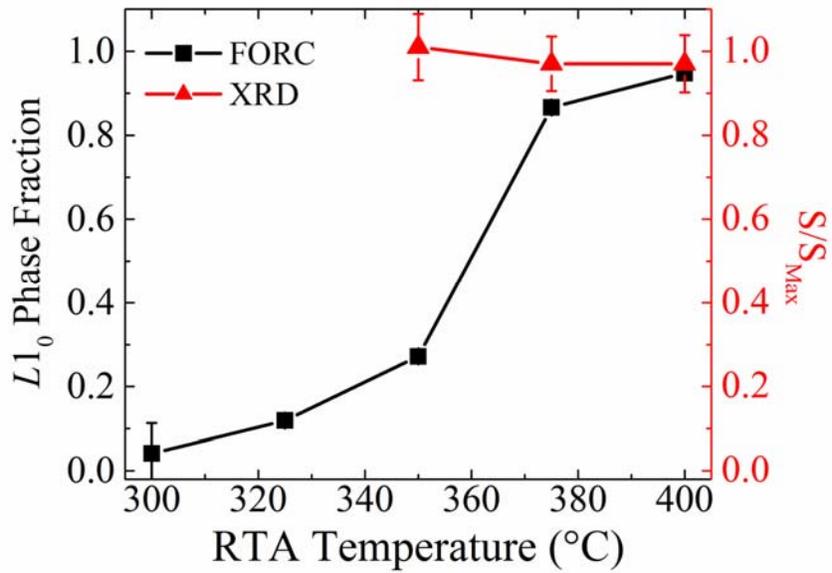

Fig. 5

# Probing the A1 to $L1_0$ Transformation in FeCuPt Using the First Order Reversal Curve Method

*Dustin A. Gilbert,*[1] *Jung-Wei Liao,*[2] *Liang-Wei Wang,*[2] *June W. Lau,*[3] *Timothy J. Klemmer,*[4] *Jan-Ulrich Thiele,*[4] *Chih-Huang Lai,*[2] *and Kai Liu*[1,*]

[1]Physics Department, University of California, Davis, CA 95616 USA

[2]Department of Materials Science and Engineering, National Tsing Hua University, Hsinchu, Taiwan

[3]National Institute of Standards and Technology, Gaithersburg, Maryland 20899, USA

[4]Seagate Technology, Fremont, CA 94538 USA

**Supplementary Online Material**

## 1. Transmission electron microscopy study of $Fe_{39}Cu_{16}Pt_{45}$

The crystalline structure of the $Fe_{39}Cu_{16}Pt_{45}$ films annealed by RTA at 325°C and 400°C was investigated directly with transmission electron microscopy (TEM) and electron diffraction, shown in Fig. S1. Both samples are polycrystalline: at $T_{RTA}$=325°C, both rounded and elongated grains are observed, with an average size of ~ 6nm; at $T_{RTA}$= 400°C, only elongated grains are observed, with an average size of ~ 12nm.

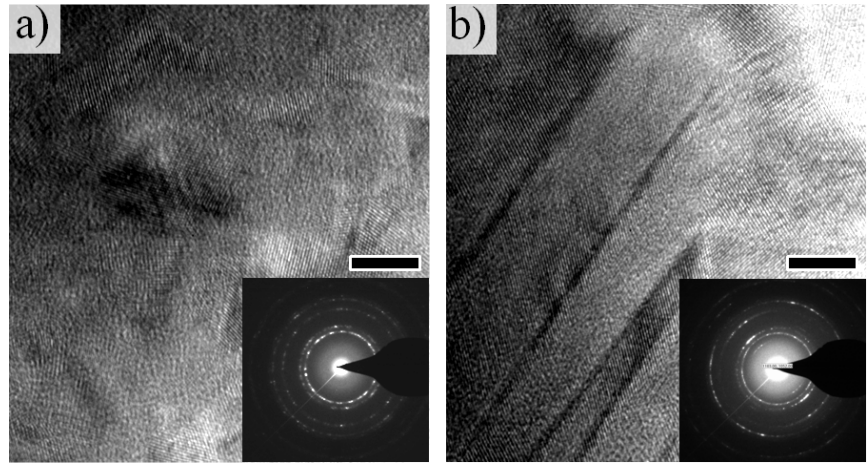

Fig. S1. Plan-view TEM images and electron diffraction patterns (insets) of $Fe_{39}Cu_{16}Pt_{45}$ films annealed by RTA at (a) 325°C and (b) 400°C. Scale bar indicates 5nm.

## 2. $Fe_{28}Cu_{27}Pt_{45}$ series

The $A1$ to $L1_0$ phase transformation was also investigated in a second series of $Fe_{28}Cu_{27}Pt_{45}$ films fabricated following the same procedure outlined in the main text and annealed between $T_{RTA}$ of 300°C and 400°C.

Similar to the $Fe_{39}Cu_{16}Pt_{45}$ films discussed in the main text, the XRD patterns did not exhibit any appreciable peaks for films annealed at $T_{RTA}$ <350°C, as shown in Fig. S2. Further, the order parameter $S/S_{Max}$ was calculated to be 0.99 for the 400°C sample. The FORC diagrams, shown in Fig. S3, again have two distinct features representative of the two magnetic phases. Similar to the main text the high-coercivity FORC feature was integrated and the phase fraction determined. The phase fractions for both series of films are plotted in Fig. S4.

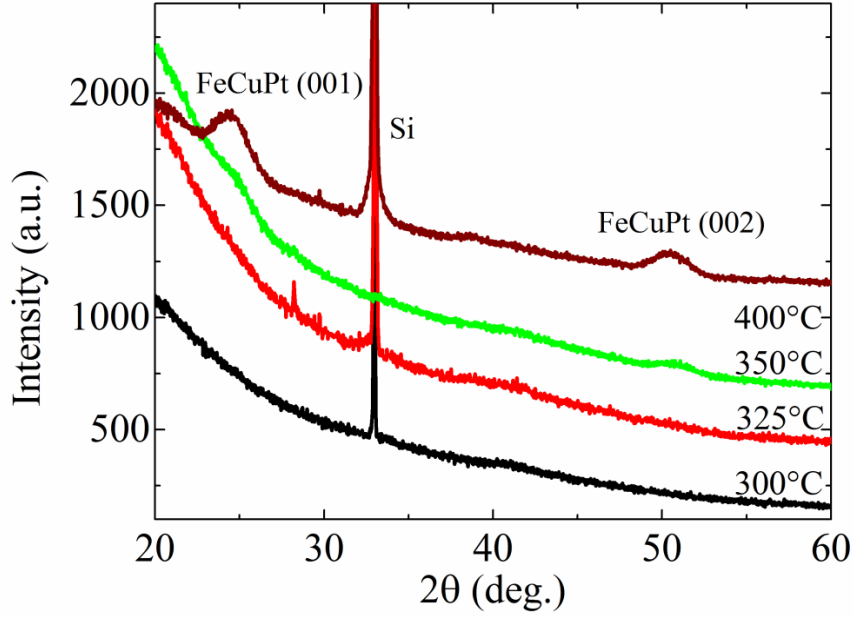

Fig. S2. X-ray diffraction patterns of $Fe_{28}Cu_{27}Pt_{45}$ thin films annealed by RTA at 300-400°C. The sharp (002) peak of the Si substrate is suppressed in some of the samples due to a slight deviation from the Bragg condition, which has no appreciable effect on the FeCuPt peaks.

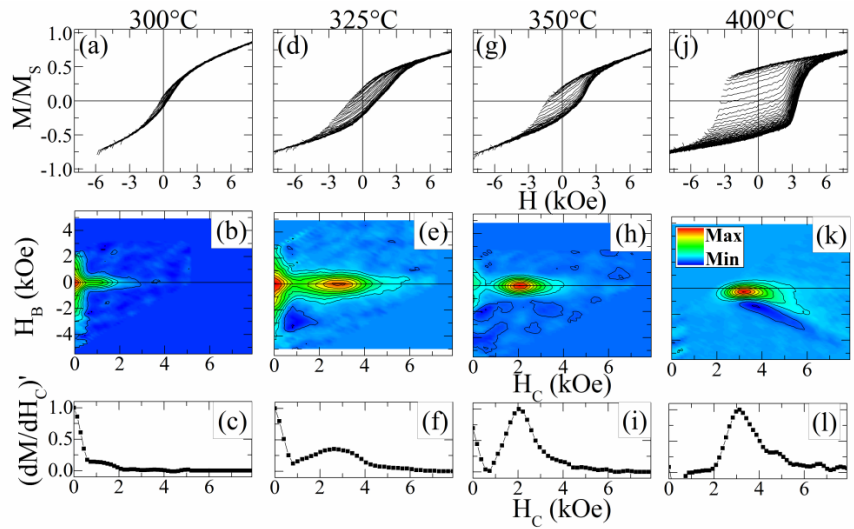

Fig. S3. Family of FORC's (top row), corresponding FORC distributions (middle row) and their normalized projections $(dM/dH_C)'$ onto the local coercivity $H_C$ axis (bottom row) for $Fe_{28}Cu_{27}Pt_{45}$ samples annealed at 300° (a-c), 325° (d-f), 350° (g-i), and 400°C (j-l).

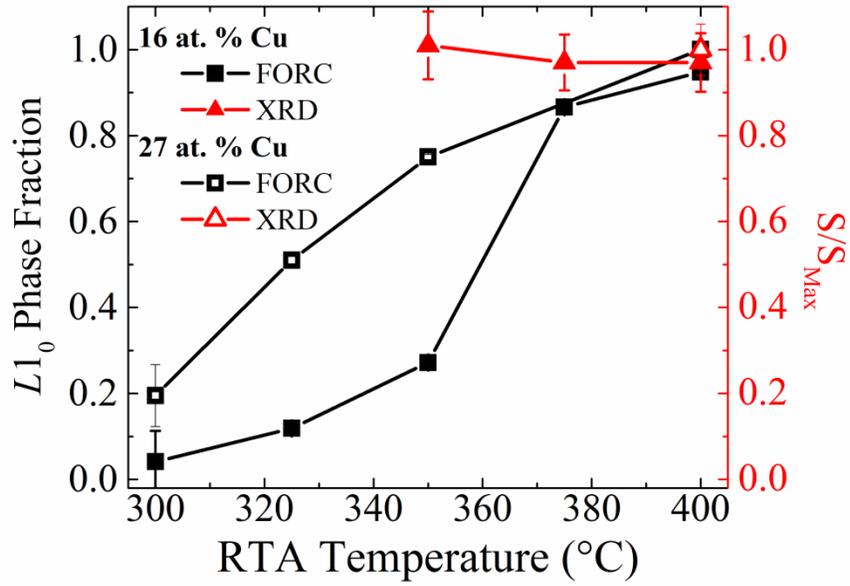

Fig. S4. $L1_0$ phase fraction determined by FORC (squares) and the order parameter determined by x-ray diffraction (triangles) for $Fe_{39}Cu_{16}Pt_{45}$ (solid symbols) and $Fe_{28}Cu_{27}Pt_{45}$ (open symbols). Error bars were determined by the error on the peak area fitting (XRD) and the propagated background standard deviation (FORC).